# Studies on the high-temperature ferroelectric transition of multiferroic hexagonal manganite $RMnO_3$

Hasung Sim[1,2], Jaehong Jeong[1,2], Haeri Kim[1,2], S-W Cheong[3], and Je-Geun Park[1,2*]

1. *Center for Correlated Electron Systems, Institute for Basic Science (IBS), Seoul 08826, Korea*
2. *Department of Physics & Astronomy, Seoul National University, Seoul 08826, Korea*
3. *Rutgers Center for Emergent Materials and Department of Physics and Astronomy, Rutgers University, Piscataway New Jersey 08854, USA*

\* Corresponding author: jgpark10@snu.ac.kr

Abstract

Hexagonal manganites are multiferroic materials with two highly-dissimilar phase transitions: a ferroelectric transition (from $P6_3/mmc$ to $P6_3cm$) at a temperature higher than 1000 K and an antiferromagnetic transition at $T_N$=65 - 130 K. Despite its critical relevance to the intriguing ferroelectric domain physics, the details of the ferroelectric transition are yet not well known to date primarily because of the ultra-high transition temperature. Using high-temperature X-ray diffraction experiments, we show that the ferroelectric transition is a single transition of abrupt order and R-$O_p$ displacement is the primary order parameter. This structural transition is then simultaneously accompanied by $MnO_5$ tilting and the subsequent development of electric polarization.

PACS Numbers: 75.85.+t, 77.80.-e, 64.70.Nd

## 1. Introduction

Multiferroic materials have attracted considerable interest over the past decade or so. Since the initial search for a new class of materials in early 2000s [1], we have witnessed a growing number of compounds joining the category of multiferroic systems. Here it is important to note that the actual history of the multiferroic studies go back much earlier to 1960s, when some of the original works were done in particular on the hexagonal manganites ($RMnO_3$) [2,3]. However, it is fair to say that among the numerous multiferroic compounds two materials stand out most. One is the rare room-temperature multiferroic $BiFeO_3$ and another is the hexagonal manganites. The hexagonal manganites do not offer as much potentials in applications as $BiFeO_3$. Nonetheless, they have been a constant source of new discoveries.

Hexagonal manganites ($RMnO_3$) with rare-earth elements of smaller ionic size (R = Sc, In, Y, Ho - Lu) have both a ferroelectric transition ($T_C$ >1000 K) and an antiferromagnetic transition ($T_N$ =65 -130K). It is also worth noting that $DyMnO_3$ can be stabilized with the hexagonal structure by growing under oxygen deficit conditions using the floating zone method. During the ferroelectric transition from a high-temperature $P6_3/mmc$ structure to a low-temperature $P6_3cm$ structure, a net electric polarization is induced, reaching a value of 5.6 $\mu C/cm^2$ at room temperature for $YMnO_3$ [4,5]. Naturally, the structural aspect of both transitions has been the focus of intensive research [6,7]. As compared to the perovskite structure with rare-earth elements of larger ionic size (R = La – Ho), the hexagonal $RMnO_3$ forms a unique structure with an ideal two-dimensional (2D) triangular lattice with $MnO_5$ polyhedron in the paraelectric phase of the $P6_3/mmc$ space group. At low temperatures, they undergo a magnetic ordering around 75 - 120 K with a so-called 120° noncollinear structure while the rare-earth moments go through their own ordering at much lower temperature, typically below 10 K. When the Mn moments order, it is accompanied by a large spin-lattice coupling [6]: which in turn appears to trigger a magnetoelectric coupling. More recently, it was shown that the noncollinear magnetic order provides an unusual magnon-phonon coupling, producing new magnetoelastic modes [7].

Although the space groups at both high and low temperatures have been rather well known since 1970s or before [8,9], the exact structural transition path between the two phases still need further investigation. In principle, based on group-subgroup analysis, there can be 4 possible paths of structural transitions: they are the 3 modes of distortion for $P6_3/mmc$: $\Gamma_2^-$ at q=(0,0,0), and $K_1$ and $K_3$ at q=(1/3,1/3,0). One additional mode is $\Gamma_1^+$ that is fully symmetric. Some paths allow an intermediate phase of different space groups while others directly go to the low-temperature $P6_3cm$ structure [10,11,12]. A complete understanding of the nature of the ferroelectric transition is extremely important, not least because the high-temperature structural transition induces the unusual trimerization of $Mn^{3+}$ ions and subsequently generates the six ferroelectric domains with intriguing vortex patterns: which itself has become the focus of intense recent studies [13].

Among the two early theoretical works on the high-temperature transition, the first one was done by van Aken et al. [5]: using a first-principle density functional theory (DFT) and analyzing their X-ray diffraction data, they came to a conclusion that 'bulking of $MnO_5$ polyhedron accompanied by Y-$O_P$ displacement' produces an energy minimum for the ferroelectric phase. About the same time, Fennie and Rabe carried out their first-principle calculations [14] independently. According to their calculations, only the $K_3$ mode is possible among the four paths with the $K_3$ mode inducing the $\Gamma_2^-$ mode and producing the electric polarization [14].

As regards the existence of the intermediate phase, there have been several conflicting experimental reports. For example, at least two papers claimed that $RMnO_3$ has a paraelectric $P6_3cm$ intermediate phase [12,15]. The reported transition temperatures were around 1200-1450 K for the structure transition and around 850-1100 K for the middle phase transition [12,15]. On the other hand, two other groups argued for a different intermediate phase of $P6_3/mcm$ [16,17]. However, Jeong et al. claimed that there is no evidence for any intermediate phase based on their total scattering data [18]. Even for the $P6_3cm$ intermediate phase, there were two different reports about its ferroelectric nature: one is that the intermediate phase should be antiferroelectric while there is also a report that the intermediate phase be ferroelectric [19].

A more recent report by Lilienblum et al. [20] suggests that the supposedly second transition to the putative intermediate phase is not a real phase transition, but rather caused by finite-size scaling effects. They measured the electric polarization by a second harmonic generation (SHG) technique and calculated the $MnO_5$ tilting mode and the discreteness parameter. Their discreteness parameter can be understood as the movement of Mn atoms, which is basically the $K_1$ mode in our case. They argued that this discreteness parameter arising from the finite size effects makes the false claim of a second transition [20].

In order to answer the question of the transient phase transition, we have undertaken careful high-temperarure x-ray diffraction studies on all the hexagonal manganites. As compared to the previous studies, the key difference in our studies is that we identified the four principal Bragg peaks, whose temperature-dependent intensity would faithfully reflect the possible structural transitions associated with the four distortion modes. Using this approach, we are able to distinguish among the roles played by each mode regarding the ferroelectric transition. We further discuss the implications of our data from the viewpoints of both experiment and theory.

## 2. Experiments

Polycrystalline $RMnO_3$ samples with R = Y, Lu-Ho were prepared by a standard solid-state reaction method with stoichiometric $R_2O_3$ and $Mn_2O_3$ of reagent grade. Hydroscopic starting materials were pre-heated at 900 °C for 24 hours before weighing the mass and the final sintering was carried out at 1300 -1400 °C for 24 hours. Since $InMnO_3$ is harder to stabilize, an extra amount of $Bi_2O_3$, at a level of few weight %, was added before the final sintering. All our samples were checked by X-ray powder diffractometer (Bruker, D8 Advance), and found to form in the single phase of the $P6_3cm$ structure, except for $InMnO_3$. $InMnO_3$ sample contains a small amount of impurity, but its impurity peaks are well separated from the main peaks of hexagonal $InMnO_3$ so we can ignore it for our analysis.

We undertook high-temperature X-ray powder diffraction experiments to examine in detail the temperature dependence of several strong Bragg peaks from 700 to 1475 K with the high-resolution diffractometer (Bruker

D8 Advance). All our measurements were carried out at every 25 K step with the counting time of 30 – 300 sec for every 0.02° step. For the (212) peak for YMnO$_3$, we reduced the temperature step further to 3 K over the temperature range between 1210 and 1260 K in order to obtain more data points with finer steps near the ferroelectric transition. Simulation and analysis were done by using FullProf [21].

3.  Results and analysis

One of the key issues was how to determine a possible intermediate phase as discussed in Refs. [16,17]. In principle, there are two possible space groups, P6$_3$mc and P6$_3$/mcm, for the intermediate phase, which has a group-subgroup relation with the P6$_3$/mmc and P6$_3$cm space groups [11]. To illustrate the key differences in the diffraction patterns for each of the four space groups, we simulated the X-ray diffraction patterns in Fig. 1(a). Here one can immediately notice that careful diffraction studies can, in principle, distinguish each phase from one another. Since the room temperature structure of P6$_3$cm has a larger unit cell than P6$_3$/mmc, there are several superlattice peaks if a direct transition should occur between the two end phases. On the other hand, if the intermediate phase is of P6$_3$mc with no Mn trimerization, there can be only an increase in the intensity of the (006) peak at 2θ=48°.

Based on the experimental data shown in Fig. 1(b), we can easily rule out both intermediate phases of either P6$_3$mc or P6$_3$/mcm as our data exhibit several superlattice peaks, notably the (212) peak, immediately below the transition temperature that cannot be explained by either of the two phases. P6$_3$/mcm phase have the same Laue class with P6$_3$cm phase, and so the (212) peak can appear but its intensity is relatively negligible as compared with the (204) peak (see the simulation in Fig. 1(b)). We also note that the (212) peak's intensity is comparable with the (204) peak's intensity immediately below transition temperature although the (212) peak is not allowed in the P6$_3$mc phase. Further close examination of the overall diffraction patterns and careful comparison with those simulation results reveal that the ferroelectric transition is a direct transition from the high-temperature P6$_3$/mmc to the room temperature P6$_3$cm, i.e. no intermediate phase with a different space group.

Hexagonal manganites undergo a structural transition from the high-temperature P6$_3$/mmc structure to the low-temperature P6$_3$cm structure, with the four possible modes: $\Gamma_1^+$, $\Gamma_2^-$, K$_1$ and K$_3$ [22]. Note that $\Gamma_1^+$ is a symmetric mode so not directly responsible for the structural transition. Fig. 2 illustrates the detailed movement of atoms for each mode. For example, the K$_3$ mode shown in Fig. 2(a) includes the MnO$_5$ tilting and the corrugation of rare-earth elements [11,12]. The MnO$_5$ tilting includes the movements of the apical oxygen atoms (O$_A$) along the ab direction, which would rotate the MnO$_5$ polyhedron. On the other hand, the corrugation of rare-earth elements includes the movement of rare-earth elements and the planar oxygen atoms (O$_p$) along the c axis.

Importantly, the apparently complex displacements involving the $K_3$ mode are antiferroelectric in nature, and so if the coupling to $\Gamma_2^-$ can be turned off the system would remain in a state with zero polarization. Fig. 2(b) also shows the $\Gamma_1^+$ breathing mode, in which $O_P$ atoms move antiferroelectrically and distort the $MnO_5$ polyhedron. For comparison, the $K_1$ mode shown in Fig. 2(c) involves $MnO_5$ movements and the Mn trimerization, which directly leads to the $P6_3/mcm$ phase. In this case, the planar oxygen atoms ($O_P$) and Mn atoms move on the ab plane, which also distorts the $MnO_5$ polyhedron. And then there is the only polar mode of $\Gamma_2^-$ in Fig. 2(d). This $\Gamma_2^-$ mode shifts all the rare-earth elements together along the c axis whereas the planar oxygen atoms ($O_P$) move in the opposite direction, producing the net electric polarization along the c axis. We note that the apical oxygen ($O_A$) atoms move in the same direction as the rare-earth elements and reduce the size of the total electric polarization.

As we have already ruled out the $K_1$-driven $P6_3/mcm$ as an intermediate phase, we can focus on the two other possibilities: $\Gamma_2^-$ and $K_3$, for our subsequent discussion. Since each atom move independently and they are in irreducible representation, we can think of separating them to trace each movements as a function of temperature. Of great importance, we can further separate the $K_3$ mode into two distinct structural motions: the R-$O_P$ displacement along the c direction and the $MnO_5$ tilting. True, these two modes are interconnected for the actual transition in real materials. However, by dividing the modes up into the two different parts of the structural change, we will be able to better understand which one of the two structural changes of the $K_3$ mode is more important, i.e. as the primary order parameter, as we will demonstrate later in the discussion.

For further discussion, several theoretical diffraction patterns are simulated in Fig. 3 for the three different cases of distortion together with the two space groups of $P6_3/mmc$ (thick line) and $P6_3cm$ (dash-dot line). To make our discussion simple and transparent, we have assumed in our simulation that one of the two distortions of the $K_3$ mode can be switched off individually. For example, we kept the $MnO_5$ tilting off in our simulation (thin line) for the only R-$O_P$ displacement distortion of the $K_3$ mode: this R-$O_P$ displacement involves the z-position movement of R and $O_p$ atoms (see also Fig. 2). For this R-$O_P$ displacement the two sites of the rare-earth atoms move in the opposite directions with a ratio of 2:1, which would accidentally cancel out the total induced dipole moment. Our simulation for the $MnO_5$ tilting (dashed line) requires the ab-plane movement of $O_A$ atoms in the $K_3$ mode. Finally, we also show our simulation results for the $\Gamma_2^-$ mode (dash-dot-dot line). For this case, R and both oxygen ($O_p$ and $O_A$) move in the same direction and induce the total polarization along the c-axis, i.e. becoming the only polar mode. We stress that this method of analyzing the diffraction patterns is rather new. As we are going to demonstrate below, it can be a quite powerful way of understanding structural transitions, whose order parameter are often difficult to discern by a more conventional Rietveld analysis.

An interesting pattern already emerges in our simulation results shown in Fig. 3 that the Bragg (211) peak is mostly due to the MnO$_5$ tilting: we estimate around 90% of the total intensity comes from the MnO$_5$ tilting distortion. For comparison, a similar percentage of the (212) and (204) peaks results from the R-O$_P$ displacement. This observation allows us to separate the relative contribution of the two distortions of the K$_3$ mode and, more importantly, to follow up the temperature dependence of each distortion individually. For the $\Gamma_2^-$ mode, we found that the (006) peak appears to follow the ferroelectric structural transition and the development of the polarization. As it becomes clear now, this possibility of being able to examine the individual distortions independently is a great advantage of our approach in this work as compared with the full Rietveld refinement analysis done in many previous reports.

With this information, we are now ready to look at the real data. We show the raw data in Fig. 4 for both YMnO$_3$ and ErMnO$_3$. As can be seen in the contour plots, all the Bragg peaks, mostly the (204) and (212) peaks, show a clear change at the reported ferroelectric transition temperature (white line). In the case of the (211) peak, it is relatively weak and so harder to follow as it gets closer to the transition temperature. For this reason, we have made two separate plots of the (211) peaks for both samples in Figs. 4c and 4d. In the blown-up pictures, the (211) peak disappears clearly above the transition temperature for both systems. On the other hand, the (006) peak of YMnO$_3$ shows a much slower temperature dependence as we will discuss later. For ErMnO$_3$ the (006) peak is located very close to the (212) peak and it is rather difficult to separate it from the other peaks. Nonetheless, our data show that all the superlattice peaks disappear at the same temperature, indicative of a single transition: which is different from some of the previous reports [19].

We show the ionic size dependence of the ferroelectric transition in Fig. 5. As RMnO$_3$ with R = Tm, Yb and Lu have their ferroelectric transition temperature above the maximum temperature of our systems, we took the data point (open circle) from Ref. [23]: the filled symbols are from our experiments reported here. As one can see, with the ionic size decreasing from Lu to Y the transition temperature gets linearly reduced. The only exception is InMnO$_3$ that shows the lowest transition temperature among all RMnO$_3$ although In has the smallest ionic radius. This difference may well be due the fact that a different structure P$\bar{3}$c was reported for InMnO$_3$ as discussed in Ref. [24].

We would now like to examine the temperature dependence of the four Bragg peaks in further detail. For this, we fitted each of the peaks using a pseudo-Voigt function by fitting the data with a fixed ratio of 35% Gaussian & 65% Lorentzian, and a fixed wavelength ratio (Cu k$_{\alpha1}$ : Cu k$_{\alpha2}$ = 2:1. We have taken great care in subtracting background off to have the correct estimate of intensity by examining the actual background signals to make it sure that there is no extra weak peak hidden under them. We plot the estimated area of the peaks as a function of temperature for YMnO$_3$ and ErMnO$_3$ in Fig. 6. This plot makes an immediately noticeable point about the temperature dependence: the (204) and (212) peaks change rather abruptly at the transition temperature while the (211) and (006) peaks are very gradual across the transition temperature.

This abrupt behavior of the (204) and (212) peaks can be taken as evidence that the phase transition may be of the weak 1st order. The lines underneath the two peaks in our Figs. 6(a) and 6(b) are our theoretical calculations using the Ginzburg-Landau analysis for a 1st order transition (solid line): $F = \frac{\alpha_2}{2}Q^2 + \frac{\alpha_4}{4}Q^4 + \frac{\alpha_6}{6}Q^6$, F is the Ginzburg-Landau free energy and Q is the displacement amplitude with $\alpha_2 = a(T_C-T)$ and $T_C$ being a critical temperature [25]: the dashed line in Fig. 6a is for what is expected of a 2nd order transition. We used the following parameters for the coefficients of the Ginzburg-Landau function: a=0.000815, $\alpha_4$ =-0.156656, and $\alpha_6$=0.064452. We used $T_C$ values of 1236 and 1421 K for YMnO$_3$ and ErMnO$_3$, respectively. Interestingly, the 3D XY model used in Ref. [13] fails to fit the temperature dependence too (dashed-dot line in Fig. 6a). For comparison, the temperature dependence of both (211) and (006) peaks can be explained by Ginzburg-Landau analysis for a 2nd order transition in Fig. 6(b). We also plotted the usual Debye-Waller (D-W) factor in Fig. 6(b). A similar plot is made for ErMnO$_3$ in Fig. 6c. A passing comment, we acknowledge that our data do not show a clear sign of thermal hysteresis although there is a better agreement with a scenario of 1st order than that of 2nd order: which is why we are arguing that we observe a 'weak' 1st order phase transition. A plausible explanation may be found in the arguments recently presented by Ref. [20] that different experimental techniques may yield seemingly contradictory results because of different coherence lengths probed by the experimental techniques.

4. Discussion

Here we would like to comment that our observation of the abrupt or weak 1st order phase transition appears not to be consistent with the estimated coefficients of the zero-temperature energy surface in the previous DFT calculations [5,14]. However, given how high the transition temperature is, it is not unreasonable to assume that the coefficients extracted from the zero-temperature DFT calculations get renormalized and the sign of $\alpha_4$ can change at higher temperature as our analysis indicates.

If we combine both the experimental data shown in Fig. 6 and the simulation result in Fig. 3, the following picture emerges of the ferroelectric transition. The R-O$_p$ displacement of the K$_3$ mode that is responsible for the (204) and (212) peaks undergoes a probably weak 1st order phase transition, and so becomes the primary order parameter of the ferroelectric transition. Simultaneously the MnO$_5$ tilting of the K$_3$ mode that is responsible for the (211) peak gets stabilized subsequently and becomes a secondary order parameter. On the other hand, there is a very gradual increase in the intensity of the (006) peak that arises from the polar $\Gamma_2^-$ mode on top of the usual

Debye-Waller factor in Fig. 6(c). Within the uncertainty of our experiment, we conclude that the polarization, i.e. the (006) peak, starts to develop right below the structural transition. To demonstrate this point further, we subtracted the theoretical temperature dependence of the Debye-Waller factor from the measured intensity of the (006) peak and show the difference in the Fig. 6(b). This is also consistent with the experimentally measured polarization by the SHG technique in Ref. [20].

We would now like to compare our main conclusion with those of previous reports. Frist, van Aken et al. discussed the relationship between the Mn-O polyhedral buckling and the Y-$O_P$ displacement [5]. In their paper, they argued that the Mn-O polyhedral buckling is a primary order parameter, and the ferroelectric displacements of Y-$O_P$ and the buckling (or tilting) of Mn-O bonds are energetically favored with respect to the centrosymmetric structure. On the other hand, Fennie et al. studied the relationship between the $K_3$ and $\Gamma_2^-$ modes and concluded that the $K_3$ phonons are strongly unstable and easily coupled to the $\Gamma_2^-$ mode [14]. They argued that the Y-$O_P$ displacement is a primary mode while the $\Gamma_2^-$ mode is secondary, in good agreement with our results. Gupta et al. carried out inelastic neutron scattering to study phonons at K and $\Gamma$ points [26], where they show that the K point phonon is unstable and leads to the ferroelectricity in agreement with our results.

However, there is also discrepancy between our results and some of the previous reports. First, in our results only the Y-$O_p$ displacement mode shows an abrupt change at the transition temperature while the $MnO_5$ tilting shows a much slower continuous temperature dependence. This observation seems to differ from what is reported, for example in the latest paper by Lilienblum et al. [20], where the $MnO_5$ tilting was claimed to have a much faster temperature dependence right below the structure transition. Then, there are several reports that the ferroelectric transition temperature is lower than the actual structure transition [10,12,27]. Unlike these reports, the temperature dependence of the (006) peak show that the actual polarization also starts right at the structural transition. We note that our conclusion on this point is consistent with those in Ref. [20]. Moreover, we would like to note that with the improper coupling of a $PQ^3$ type the phase transition of P, i.e. the intensity of (006), should be of 1st order. However, as discussed by Fennie and Rabe [14], the $PQ^3$ coupling tends to suppress P close to the transition (while P ~Q for large values of Q).

Finally, we note that our conclusion is similar to that of Ref. [19] in that there is no intermediate phase with space group $P6_3/mcm$. However, there is also discrepancy, namely the evidence of the second transition. In our data, we failed to observe any evidence for this second transition: this second transition was rather weak even in Ref. [19]. It is also worth noting that their analysis of mode-decomposition using AMPLIMODES reveal that all four modes seem to undergo rather abrupt change at the ferroelectric transition. What we have shown in this paper is that by measuring and examining each individual Bragg peaks associated with the four modes we can clearly follow the temperature dependence of each mode directly from the measured data and reach a conclusion on the different temperature dependence of the modes.

## 5. Summary

To summarize, we conclude based on the experimental data and the theoretical simulation that first, there is only a single transition of weak first order nature from the high-temperature $P6_3/mmc$ structure to the low-temperature $P6_3cm$ structure with no intermediate phase. By scrutinizing the individual distortions independently based on a new method of analysis combined with a careful modes examination, we also found that this transition is mainly driven by the R-$O_P$ displacement, which is then accompanied by the MnO$_5$ tilting mode and the polarization mode that vary more slowly. Therefore, we conclude that the R-$O_p$ displacement of the $K_3$ mode is the primary order parameter of the high-temperature structural transition.

**Acknowledgements**

We thank Cristian D. Batista, Craig J. Fennie and Yukio Noda for useful discussions and comments on the manuscript. The work at the IBS CCES was supported by the research program of Institute for Basic Science (IBS-R009-G1). The work at Rutgers University was supported by the DOE under Grant Number DOE: DE-FG02-07ER46382.

**References**

[1] N. Spaldin, 2000 *J. Phys. Chem*. **B104** 6694

[2] H. Yakel, W. Koehler, E. Bertaut, and F. Forrat, 1963 *Acta Crystallogr.* **16,** 957

[3] G. Smolenskii, V. Bokov, V. Isupov, N. Krainik and G. Nedlin 1968. *Helv. Phys. Acta,* **41,** 1187.

[4] S. H. Kim, S. H. Lee, T. H. Kim, T. Zyung, Y. H. Jeong, and M. S. Jang, 2000 *Cryst. Res. Technol.* **35** 19

[5] B. B. V. Aken, T. T. M. Plastra, A. Filippetti, and N. A. Spaldin, 2004 *Nat. Mater*. **3** 164

[6] S. Lee, A. Pirogov, M. Kang, K.-H. Jang, M. Yonemura, T. Kamiyama, S.-W. Cheong, F. Gozzo, N. Shin, H. Kimura, Y. Noda, and J.-G. Park, 2008 *Nature* **451** 14

[7] J. Oh, M. D. Le, H.-H. Nahm, H. Sim, J. Jeong, T. G. Perring, H. Woo, K. Nakajima, S. Ohira-Kawamura, Z. Yamani, Y. Yoshida, H. Eisaki, S.-W. Cheong, A.L. Chernyshev and J.-G. Park 2016 *Nat. Comm.* **7** 13146

[8] K. Łukaszewicz and J. Karut-Kaliciṅska, 1974 *Ferroelectrics* **7** 81

[9] E. F. Bertaut and M. Merciier, 1963 *Phys. Let*t **5** 1

[10] H. Sim, J. Oh, J. Jeong, M. D. Le, and J.-G. Park, 2016 *Acta Cryst*. **B72** 820

[11] H. T. Stokes and D. M. Hatch, Isotropy Subgroups of the 230 Crystallographic Space Groups, World Scientific, Singapore

[12] T. Lonkai, D. G. Tomuta, U. Amann, J. Ihringer, R. W. A. Hendrikx, D. M. Tobbens, and J. A. Mydosh, 2004 *Phys. Rev. B* **69** 134108

[13] S.-Z. Lin *et al* 2014 *Nature Phys.* **10** 970

[14] C. J. Fennie and K. M. Rabe, 2005 *Phys. Rev. B* **72** 100103(R)

[15] J. Kim, K. C. Cho, Y. M. Koo, K. P. Hong, and N. Shin, 2009 *Appl. Phys. Lett.* **95** 132901

[16] G. Nenert, M. Pollet, S. Marinel, G. R. Blake, A. Meetsma, and T. T. M. Palstra, 2007 *J. Phys. Condes. Matter* **19** 466212

[17] S. C. Abrahams, 2009 *Acta Cryst.* **B65** 450

[18] I.-K. Jeong, N. Hur, and T. Proffen, 2007 *J. Appl. Phys.* **40** 730

[19] A. S. Gibbs, K. S. Knight, and P. Lightfoot, 2011 *Phys. Rev. B* **83** 094111

[20] M. Lilienblum, T. Lottermoser, S. Manz, S. M. Selbach, A. Cano, and M. Fiebig, 2015 *Nature Phys.* **11** 1070

[21] J. Rodriguez-Carvajal, 1993 *Physica B* **192** 55

[22] D. Orobengoa, C. Capillas, M. I. Aroyo, and J. M. Perez-Mato, 2009 *J. Appl. Cryst.* **42** 820

[23] S. C. Chae, N. Lee, Y. Horibe, M. Tanimura, S. Mori, B. Gao, S. Carr, and S.-W. Cheong, 2012 *Phys. Rev. Lett.* **108** 167603

[24] Y. Kumagai, A. A. Belik, M. Lilienblum, Naëmi Leo, M. Fiebig, and N. A. Spaldin 2012 *Phys. Rev. B* **85** 174422

[25] H. Sim, D. C. Peets, S. Lee, S. Lee, T. Kamiyama, K. Ikeda, T. Otomo, S.-W. Cheong, and J.-G. Park, 2014 *Phys. Rev. B*, **90**, 214438

[26] M. K. Gupta, R. Mittal, M. Zbiri, N. Sharma, S. Rols, H. Schober and S. L. Chaplot 2015 *J. Mater. Chem. C*, **3,** 11717

[27] S. C. Abrahams, 2001 *Acta Cryst.* **B57**, 485

**Figures**

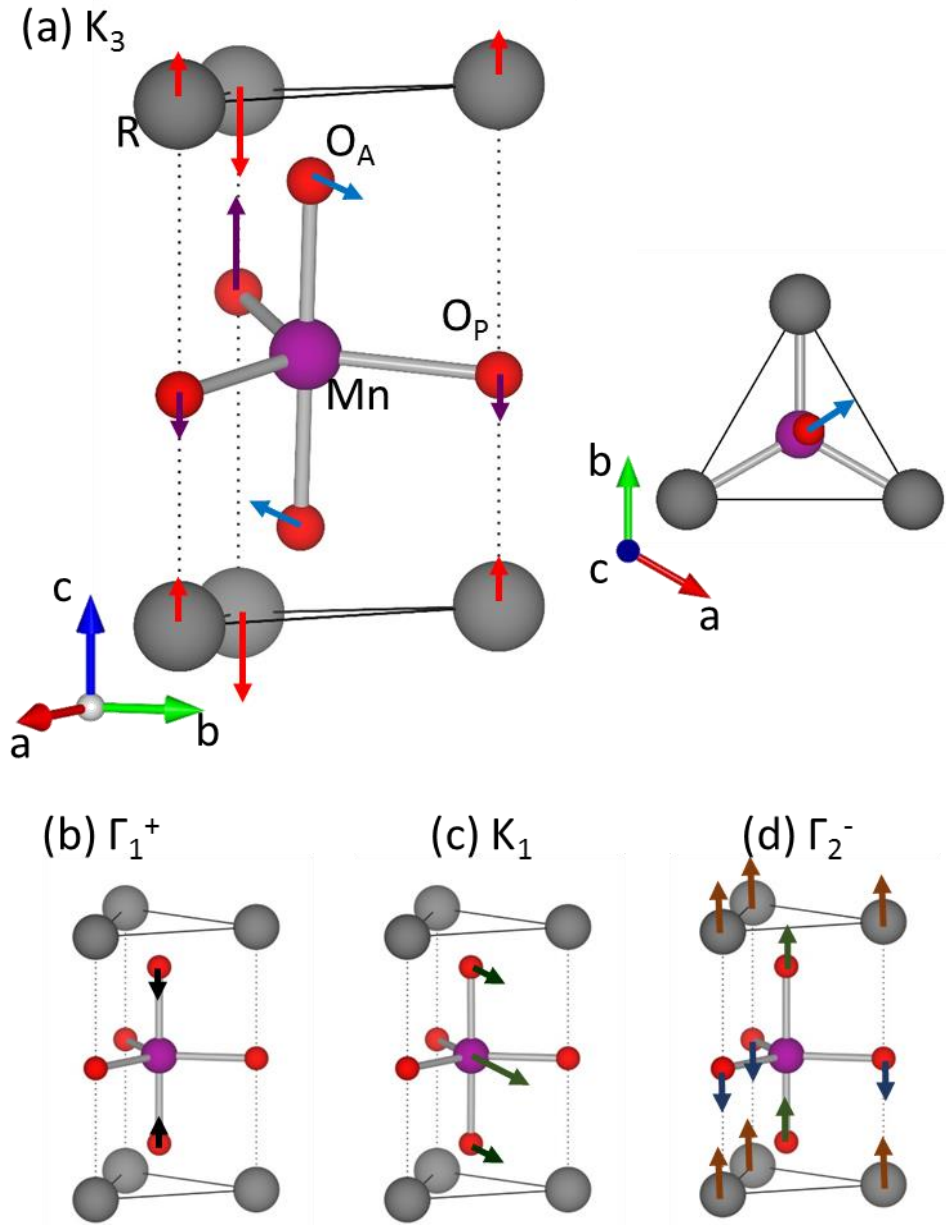

Figure 1 (color online) (a) Simulated X-ray diffraction results of YMnO$_3$ for four space groups: P6$_3$/mmc, P6$_3$/mcm, P6$_3$mc, and P6$_3$cm structures with * marking the superlattice peaks. Inset figure shows an enlarged region of the data with the (211), (204), and (212) peaks marked from left. (b) Experimental data of YMnO$_3$ collected at seven representative temperatures below and above the transition temperature. The dashed line indicates the ferroelectric transition temperature.

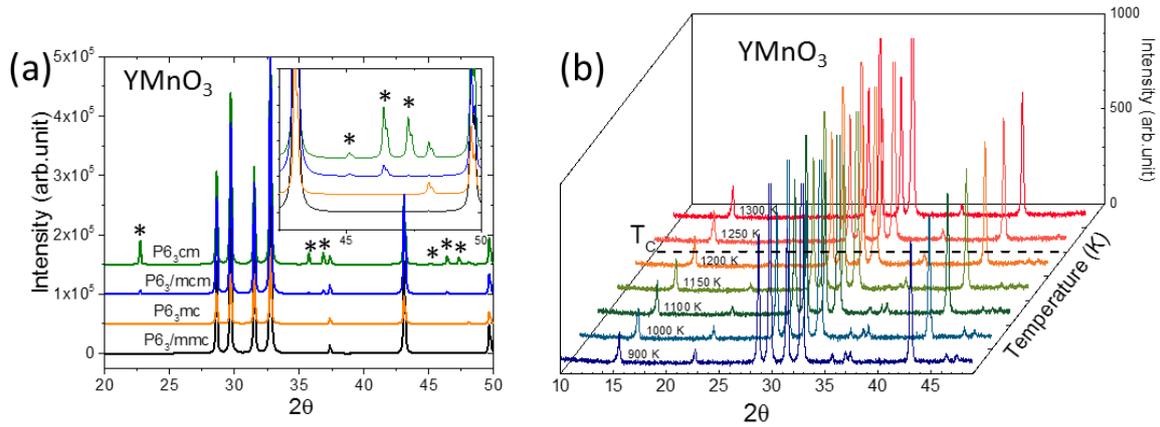

Figure 2 (color online) Direction and relative size of atomic displacements of RMnO$_3$ for the four possible modes. (a) K$_3$, (b) $\Gamma_1^+$, (c) K$_1$ and (d) $\Gamma_2^-$. The arrows in the figure indicate the direction of the atomic displacement when the corresponding mode is active, and the length of the arrows is chosen to be proportional to the magnitude of the atomic displacements.

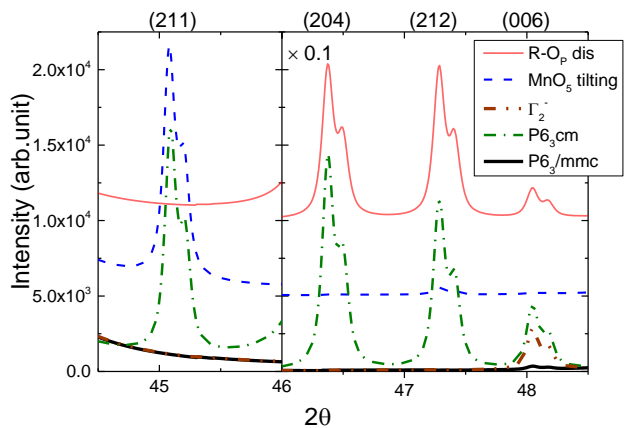

Figure 3 (color online) Simulated X-ray diffraction results on YMnO$_3$. Blown up picture of the simulation results with three specific distortions: R-O$_P$ displacement (thin line), MnO$_5$ tilting (dashed line), both of the K$_3$ mode, and the $\Gamma_2^-$ mode (dash-dot-dot line). For comparison, we also calculate the diffraction patterns for the high-temperature P6$_3$/mmc (thick line) and the low-temperature P6$_3$cm structure (dash-dot line). (HKL) lists are based on the P6$_3$cm structure.

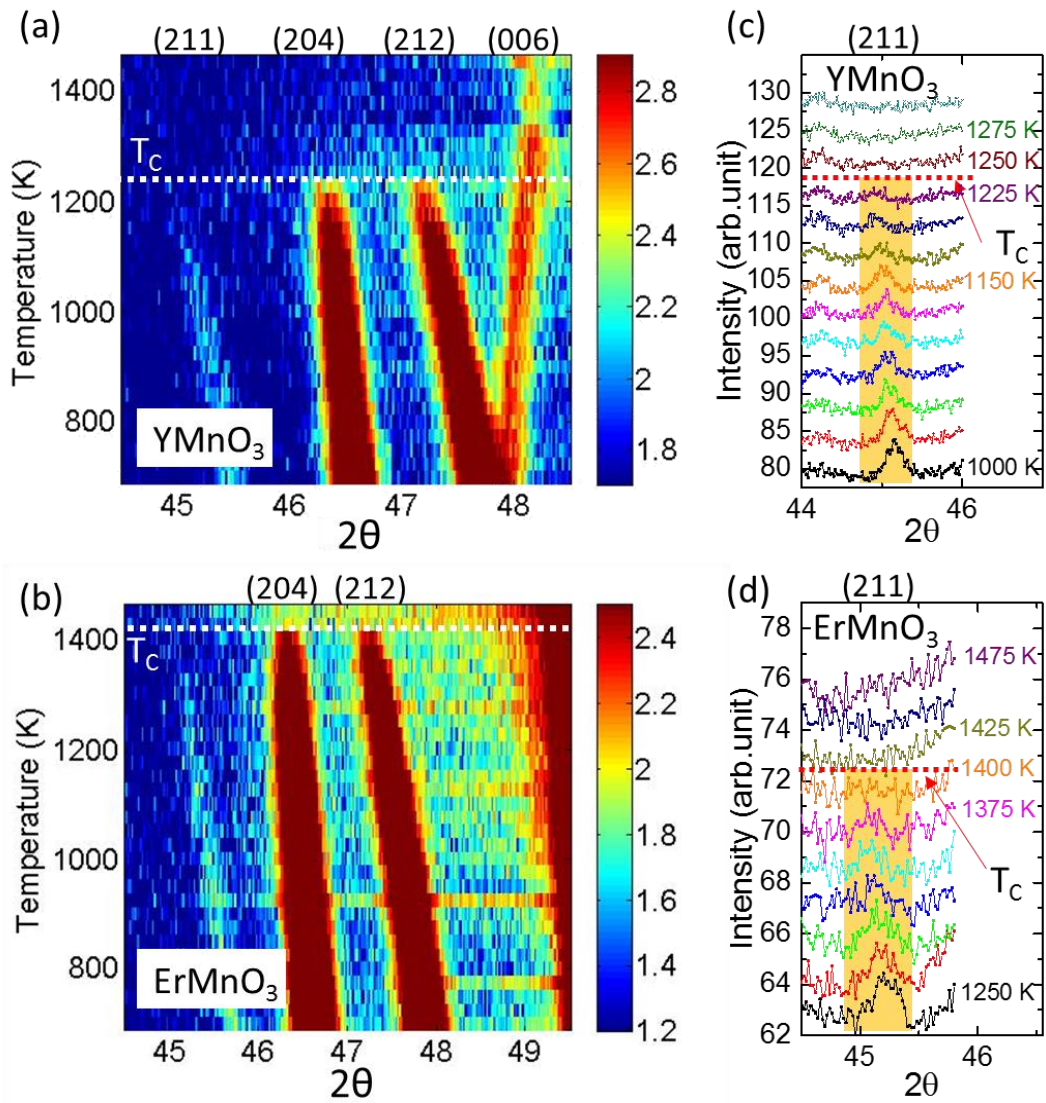

Figure 4 (color online) Temperature-dependent X-ray diffraction data. Contour plot of temperature-dependent X-ray diffraction data for (a) YMnO$_3$ and (b) ErMnO$_3$ with the transition marked by white horizontal lines. Blown-up pictures of the (211) peak are shown for (c) YMnO$_3$ and (d) ErMnO$_3$ with longer counting time.

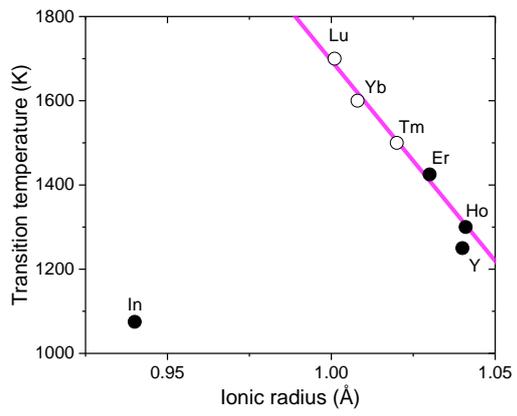

Figure 5 (color online) Ionic size dependence of the transition temperature with open circle for the data points taken from Ref. [23].

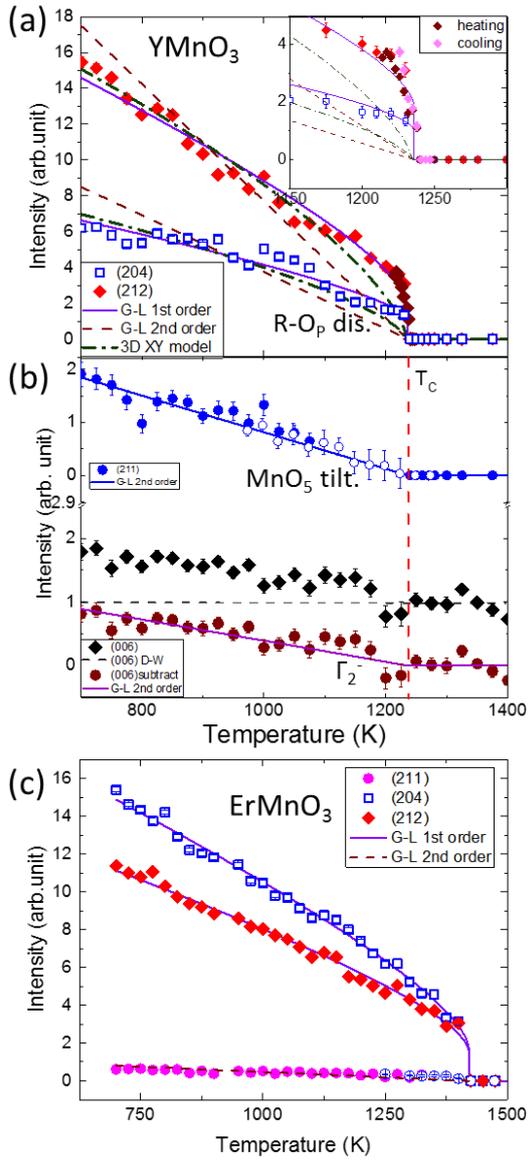

Figure 6 (color online) Temperature dependence of the peak intensity of the four Bragg peaks. (211), (204), (212) and (006) for (a, b) YMnO$_3$ and (c) ErMnO$_3$ with the lines for Ginzburg-Landau fitting results (see the text). (a) Data points were collected with much longer counting time near the transition with warming and cooling. Solid (dashed) lines in (a) show the 1$^{st}$ (2$^{nd}$) order G-L fitting results: for comparison, a line (dashed-dot) is also added for the 3D XY model used in Ref. [11]. Inset shows the blown-up picture of the data points and the fitting results near the transition temperature. (b) Circles (closed and open) represent data points of the (211) peak with the open symbols collected with longer counting time. We also made similar plots for the (006) peaks before (diamond) and after (circle) correcting for the usual temperature dependence of the Debye-Waller factor (dashed line). (c) A similar plot is made for ErMnO$_3$